\documentclass[a4paper]{article}
\usepackage[margin=24mm]{geometry}

\usepackage{amssymb}
\usepackage{amsmath}
\usepackage{graphicx}
\usepackage{color}

\newcommand{\be}{\begin{equation}}
\newcommand{\ee}{\end{equation}}
\newcommand{\bea}{\begin{eqnarray}}
\newcommand{\eea}{\end{eqnarray}}
\newcommand{\de}{\delta}
\newcommand{\bk}{\mathbf{k}}

\newcommand{\lftwo}{{\sc lat}field{\sc 2}\ }

\begin{document}

\title{General relativity and cosmic structure formation}

\author{Julian Adamek$^1$, David Daverio$^1$, Ruth Durrer$^1$ \& Martin Kunz$^{1,2}$}

\date{\vspace{-20pt}}

\maketitle

\begin{enumerate}
 \item[${}^1$] D\'epartement de Physique Th\'eorique \& Center for Astroparticle Physics, Universit\'e de Gen\`eve,\\24 Quai E.\ Ansermet, 1211 Gen\`eve 4, Switzerland
 \item[${}^2$] African Institute for Mathematical Sciences, 6 Melrose Road, Muizenberg 7945, South Africa
\end{enumerate}

\vspace{5pt}

\centerline{\date{\today}}

\vspace{5pt}

\begin{abstract}
Numerical simulations are a versatile tool providing insight into the complicated process of structure
formation in cosmology \cite{Springel:2005nw}. This process is mainly governed by gravity, which is the dominant
force on large scales. To date, a century after the formulation of general relativity \cite{Einstein:1915ca},
numerical codes for structure formation still employ Newton's law of gravitation. This approximation relies
on the two assumptions that gravitational fields are weak and that they are only sourced by non-relativistic
matter. While the former appears well justified on cosmological scales, the latter imposes restrictions
on the nature of the ``dark'' components of the Universe (dark matter and dark energy) which are,
however, poorly understood.
Here we present the first simulations of cosmic structure formation using equations consistently derived from
general relativity. We study in detail the small relativistic effects for a standard
$\Lambda$CDM cosmology which cannot be obtained within a purely Newtonian framework. Our
particle-mesh N-body code computes all six degrees of freedom of the metric and consistently solves the geodesic equation for particles,
taking into account the relativistic potentials and the frame-dragging force. This conceptually clean approach is
very general and can be applied to various settings where the Newtonian
approximation fails or becomes inaccurate, ranging from simulations of models with dynamical dark energy \cite{Noller:2013wca}
or warm/hot dark matter \cite{Costanzi:2013bha} to core collapse supernova explosions \cite{Kotake:2011yv}.
\end{abstract}

\vspace{5pt}

The applicability of Newton's law of gravitation in the context of cosmic structure formation has been discussed
extensively in the recent literature \cite{Chisari:2011iq,Green:2011wc,Rigopoulos:2013nda}. In particular, it is now well
understood that this simplified description is quite accurate when applied within standard $\Lambda$CDM cosmology
where perturbations come entirely from non-relativistic matter. However, the situation is not
satisfactory for two reasons. 
Firstly, the quality  of observational data is rapidly increasing, and upcoming galaxy surveys will eventually
reach a precision where a naive treatment of the effects of spacetime geometry becomes insufficient \cite{Yoo:2014sfa,Bonvin:2014owa}.
Secondly, the true nature of dark matter and dark
energy is not yet established. In order to study models beyond $\Lambda$CDM, some of which may feature relativistic
sources of stress-energy (e.g.\ warm dark matter or dynamical dark energy), employing the Newtonian approximation
is not always justified. A number of numerical codes has been developed for particular models \cite{Schmidt:2009sg,Li:2011vk,Puchwein:2013lza,Llinares:2013jza},
yet a general framework would be desirable. Furthermore, Newtonian gravity is
acausal and not sensitive to the presence of a cosmological horizon. Even if a judicious interpretation of the output of Newtonian
simulations can cure this problem at the linear level, it comes back when one goes beyond linear perturbation theory and it would
be preferable to use the correct physics from the outset.

Moving from the absolute space and time of the Newtonian picture
towards a general relativistic view where geometry is dynamical poses
a significant conceptual challenge. Recent progress is owed to a suitable formulation of the relativistic setting in
terms of a weak-field expansion which is well adapted for (but not restricted to) cosmological applications \cite{Green:2011wc,Adamek:2013wja,Adamek:2014xba}.
Based on these ideas we have developed a numerical code, \texttt{gevolution}, which is designed to perform cosmological
N-body simulations fully in the context of general relativity, evolving all six metric degrees of freedom. 
In brief, our approach can be summarized as follows.
\begin{enumerate}
 \item We choose a suitable ansatz for the metric which is split into background and perturbations. We work in Poisson gauge
 where the perturbed Friedmann-Lema\^itre-Robertson-Walker metric is 
\begin{equation}\label{e:metric}
  ds^2 = a^2(\tau) \left[-\left(1 + 2 \Psi\right) d\tau^2 - 2 B_i dx^i d\tau + \left(1 - 2 \Phi\right) \delta_{ij} dx^i dx^j + h_{ij} dx^i dx^j\right]\, ,
\end{equation}
where $a$ denotes the scale factor of the background, $x^i$ are comoving coordinates on the
spacelike hypersurfaces, and $\tau$ is conformal time. $B_i$ (also denoted as $\mathbf{B}$) is
transverse and is responsible for frame dragging; $h_{ij}$ is transverse and traceless, it contains the two
spin-2 degrees of freedom of gravitational waves. 
 \item We assume that the perturbations of the metric, but not necessarily their derivatives, remain small on the scales
 of interest. This is a valid approximation for cosmological scales even when perturbations in the stress-energy tensor are large. 
 \item Einstein's equations are then expanded in terms of the metric perturbations but without 
 a perturbative treatment of the stress-energy tensor.
 We include all terms linear in metric perturbations and go to quadratic order in terms which have two spatial derivatives
 acting on the scalar metric perturbations $\Phi$, $\Psi$. This weak-field expansion contains
 all six components of the metric correctly at leading order.
 To determine the evolution of the metric, we numerically solve the ``$00$'' and the traceless part of the ``$ij$'' Einstein equations:
 \begin{subequations}
\begin{equation}\label{e:modPois}
 \left(1 + 4 \Phi\right) \Delta \Phi - 3 \mathcal{H} \Phi' - 3 \mathcal{H}^2 \Psi + \frac{3}{2} \left(\nabla \Phi\right)^2 = -4 \pi G a^2 \delta T^0_0\, ,
\end{equation}
\begin{multline}
 \left(\delta^i_k \delta^j_l - \frac{1}{3} \delta_{kl} \delta^{ij}\right) \biggl[\frac{1}{2} h_{ij}'' + \mathcal{H} h_{ij}' - \frac{1}{2} \Delta h_{ij} + B_{(i,j)}' + 2 \mathcal{H} B_{(i,j)} + \left(\Phi - \Psi\right)_{,ij}\biggr.\\
 \biggl. + \left(\Phi - \Psi\right)_{,i} \left(\Phi - \Psi\right)_{,j} + 2 \left(\Phi - \Psi\right) \left(\Phi - \Psi\right)_{,ij} - 2 \left(\Phi - \Psi\right) \Phi_{,ij} + 2 \Phi_{,i} \Phi_{,j} + 4 \Phi \Phi_{,ij}\biggr]\\
 = 8 \pi G a^2 \left(\delta_{ik} T^i_l - \frac{1}{3} \delta_{kl} T^i_i\right)\, .
 \label{e:ij}
\end{multline}
\end{subequations}
Eq.~(\ref{e:modPois}) is the generalisation of the Newtonian Poisson equation ($\Delta \Phi = 4 \pi G a^2 \delta\rho$) and contains additional relativistic terms. Here $\Delta \Phi$ and $\delta T^0_0 = -\delta\rho = T^0_0 - \bar{T}^0_0$ are not required to be small --
 in fact, they become much larger than the background value $\bar{T}^0_0$ inside dense regions.  Eq.\ (\ref{e:ij}) can be used to evolve all the non-Newtonian degrees of freedom of the metric: $\Phi-\Psi$, the two spin-1 degrees of freedom, $\mathbf{B}$, and the two spin-2 helicities, $h_{ij}$. To do this, we decompose
the equation into scalar, vector and tensor parts in Fourier space. 
 \item The metric is then used to solve the equations of motion for matter (and possibly other degrees of freedom);
 collisionless particles propagate along geodesics in the perturbed geometry. This determines the evolution of the stress-energy tensor.
\end{enumerate}

The implementation in \texttt{gevolution} adopts a particle-mesh scheme where the metric field is represented on a regular cubic
lattice while matter takes the form of an N-body ensemble of particles that samples the six-dimensional phase space. The stress-energy
tensor on the lattice is obtained by a particle-to-mesh projection and is used in Einstein's equations
to solve for the metric perturbations. Finally, the particles are evolved by interpolating the metric to the particle
positions and integrating the geodesic equation. \texttt{gevolution} is built on top of the \lftwo c++ framework \cite{David:2015eya}.
\lftwo handles the lattice, the fields, the parallelization (using MPI) and the Fast Fourier Transforms which \texttt{gevolution} needs for solving Eqs.\ (\ref{e:modPois}) and (\ref{e:ij}).
We have extended the public version of \lftwo to also handle the N-body particles, including projection and interpolation methods.
We plan to make these features publicly available in the near future, together with a release of the \texttt{gevolution} code.

As a first application we choose standard $\Lambda$CDM cosmology. We expect only small effects in this case, but we can use these simulations to gain
confidence in our new approach. We generate \cite{Behroozi:2011ju} two halo catalogs containing $\sim 500,\!000$ halos each, one
using our relativistic approach and one with the traditional Newtonian approach as reference, starting from identical linear initial conditions. A Kolmogorov-Smirnov test shows no significant
disagreement in the distributions of some $25$ different halo properties such as mass, spin or shape parameters.

\begin{figure}
\includegraphics[width=0.7\textwidth]{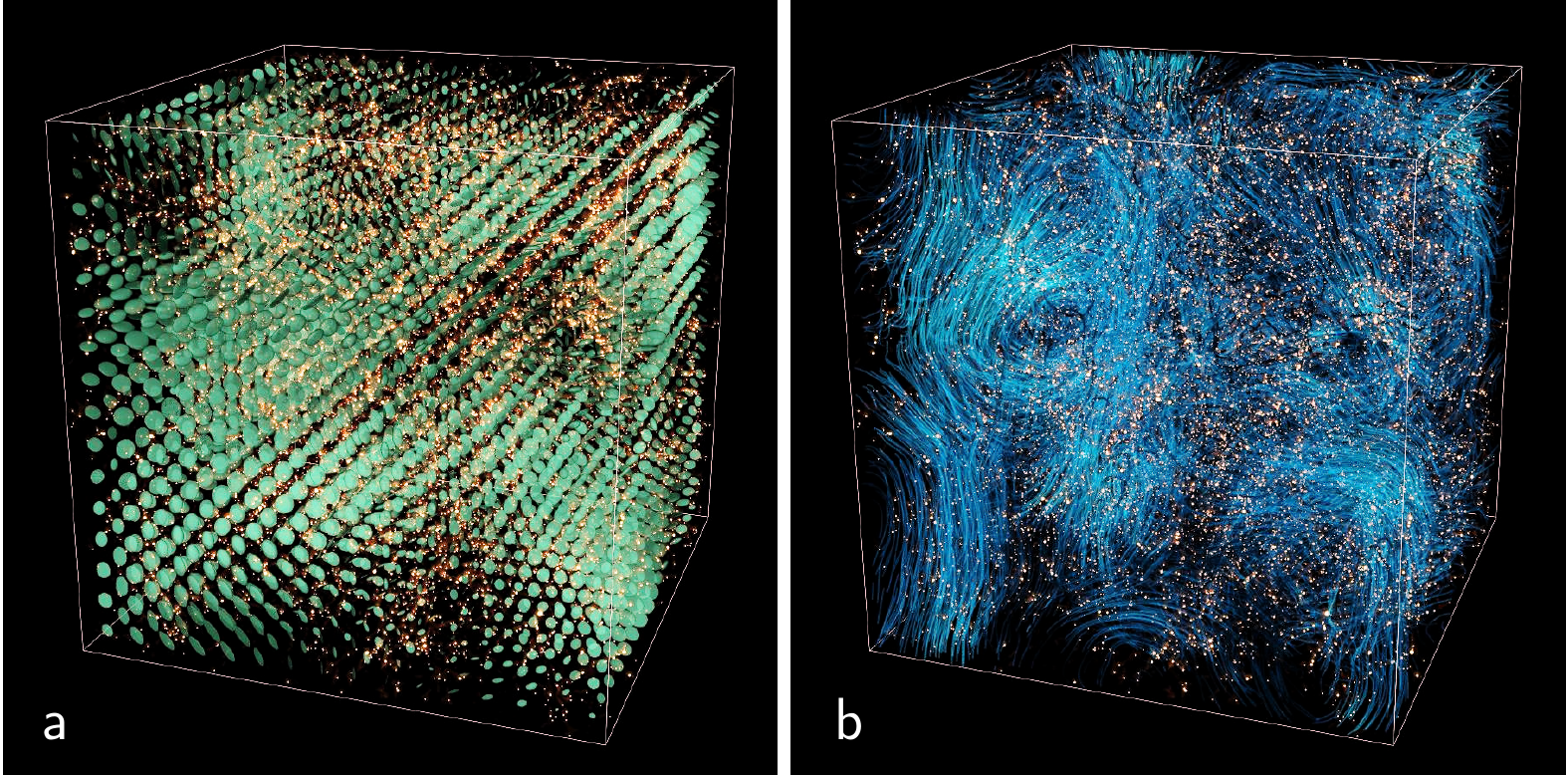}

\caption{\label{f:Box} \small \textbf{Spin-1 and spin-2 metric perturbations.} Visualisation of a simulation volume of $(512 \mathrm{Mpc}/h)^3$ at redshift $z=0$.
Dark matter halos are rendered as orange blobs scaled to the virial radius.
Panel \textsf{a} illustrates the spin-2 perturbation $h_{ij}$ by applying it as affine transformation on spheroids of fixed radius;
the shape and size of the resulting ellipsoids therefore indicates,
respectively, the polarization and amplitude of the tensor perturbation.
Panel \textsf{b} shows a stream plot of the spin-1 perturbation $\mathbf{B}$, indicating how ``spacetime is dragged around'' by vortical matter flows.}
\end{figure}

The largest non-Newtonian effect  
is frame dragging, which is associated with the spin-1 perturbation $\mathbf{B}$.
This perturbation is sourced by the curl part of the momentum density found, e.g.,
in rotating massive objects. As long as perturbations are small this is a second-order effect which has been
studied using perturbation theory \cite{Lu:2008ju}. The power spectrum for $\mathbf{B}$ has also been computed in the
non-perturbative regime of structure formation using a post-Newtonian framework \cite{Bruni:2013mua} which shares some
features with our approach. These results are useful benchmarks for our code, but we now go beyond: our
simulations track the full three-dimensional realisation of $\mathbf{B}$ (see Fig.~\ref{f:Box}). We can therefore measure the actual frame
dragging force on individual particles.

While the typical gravitational acceleration on Mpc scales is of the order of $10^{-9}$ cm/$\mathrm{s}^2$, frame
dragging contributes only at the level of $10^{-14}$ cm/$\mathrm{s}^2$ (both numbers are mass-weighted rms values
from our simulations at redshift zero), and the highest value we measure is some $10^{-12}$ cm/$\mathrm{s}^2$. Thus, for
objects moving at $1000$ km/s (a typical peculiar velocity at those scales) the $\Delta v$ due to frame dragging is no
more than $1$ km/s. It should be noted that these numbers are scale dependent, as the acceleration is larger on smaller 
scales. 

We compare our simulation results also with predictions \cite{Ananda:2006af,Baumann:2007zm} from second order perturbation theory
for the power spectra of $h_{ij}$, of $\mathbf B$ and of $\Phi-\Psi$. The latter requires regularization in the infrared,
which is implemented by the finite volume of the simulation.

In order to fully capture the amplitude of the non-Newtonian terms, it is important that the scale of
matter-radiation equality is represented in the simulation volume. On the other hand, these terms are generated by non-linearities and are
therefore amplified at small scales. In order to obtain our results we had to cover at least three orders of magnitude in scale. 
Our largest simulation used a lattice with $4096^3$ points, corresponding to
$6.7\times 10^{10}$ particles as we always start with one particle per grid point.

\begin{figure}[h]
\includegraphics[width=0.8\textwidth]{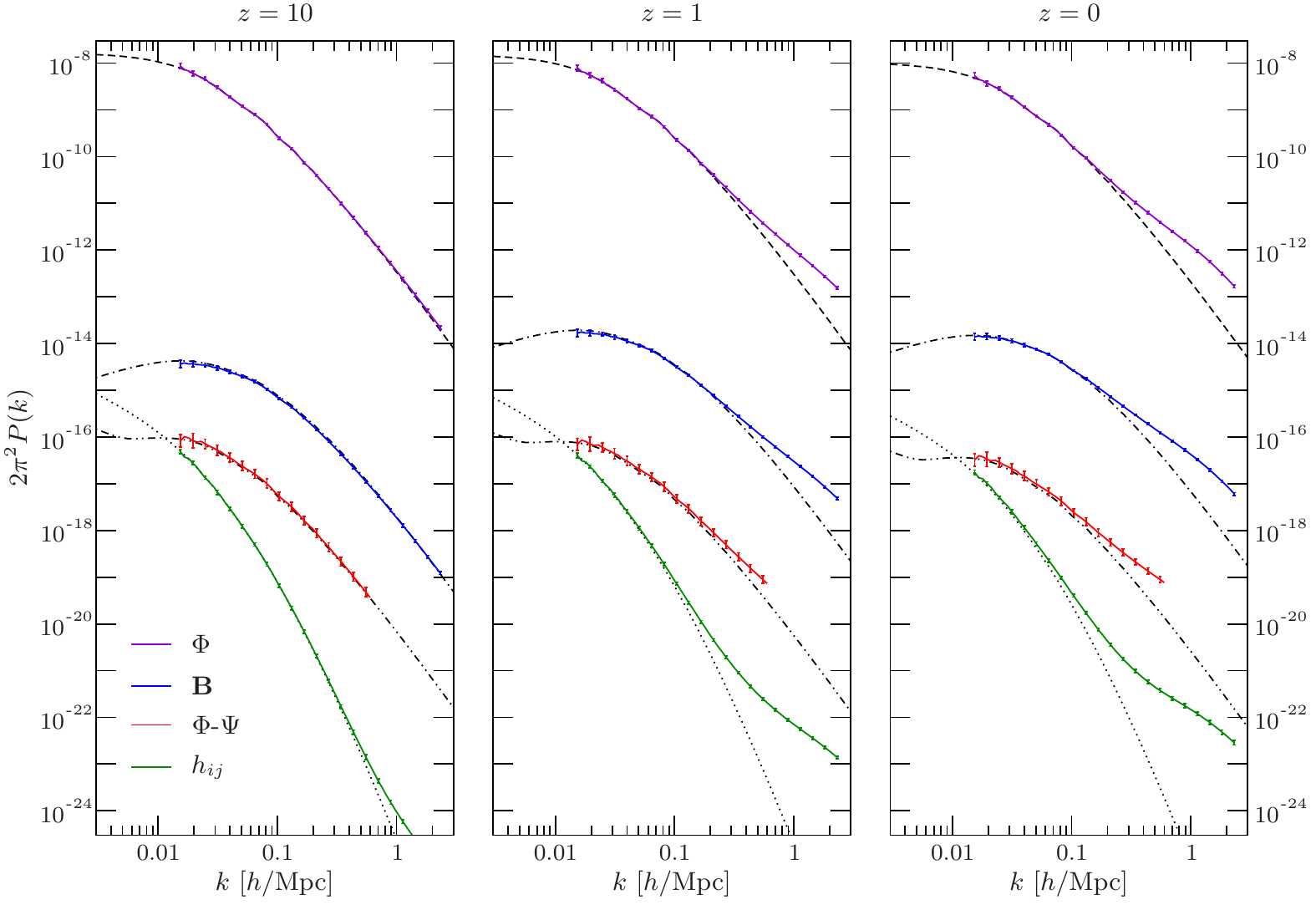}

\caption{\label{f:Pspec} \small \textbf{Power spectra.} We show the power spectra of the gravitational potential $\Phi$ (top, violet),
the frame dragging potential $\mathbf{B}$ (second from top, blue), the difference of the
relativistic potentials $\Phi-\Psi$ (third from top, red) and the tensor perturbation $h_{ij}$ (bottom, green) at redshifts $z=10, ~ 1, ~0$. The black lines (dashed, dot-dashed,
dot-dot-dashed, and dotted) indicate the corresponding results from second-order perturbation theory. As expected, the numerical
results deviate from these perturbative extrapolations at low redshift and small scales. All perturbation variables are significantly
enhanced on scales $k\gtrsim 1h/$Mpc at redshift $z=1$ and below. Error bars indicate
the random fluctuations from different realisations.
}
\end{figure}

Fig.~\ref{f:Pspec} shows the numerical power spectra at three redshifts. For these spectra we
used eight simulations with lattices of $2048^3$ points and two different box sizes, all starting at redshift $z=100$. Four simulations had a box size of ($2048$ Mpc$/h$)$^3$, the other four were ($512$ Mpc$/h$)$^3$.
In the regime where it can be trusted we find excellent agreement with second order perturbation theory
which demonstrates that our code produces valid results.
At late times and on small scales, perturbation theory breaks down.
Nonlinearities enhance the frame dragging by more than an order of magnitude at redshifts $z=1$ and $z=0$ and on scales $k\gtrsim 1h/$Mpc. The enhancement is even more dramatic for tensor perturbations.

It should finally be noted that our framework jointly solves for background and perturbations in a self-consistent way. Therefore we
confirm that clustering only has a small effect on the expansion rate of the Universe. The non-Newtonian effects which we have quantified
in the nonlinear regime of structure formation remain small within a
$\Lambda$CDM Universe, but they may nevertheless be measurable in the future.
For the first time, general relativity has been implemented as the theory of gravity in a cosmological N-body code, making
it possible to feed these effects back into the dynamics.
Our numerical framework will be particularly useful in scenarios where relativistic sources
are present, like in models of dynamical dark energy, topological defects, or with relativistic particles such as neutrinos or warm dark matter.
Such scenarios are expected to display larger relativistic effects. Contrary
to Newtonian schemes we are also able to solve the full geodesic equation for arbitrary velocities, allowing
for a realistic propagation of radiation or high-velocity particles.

\subsection*{Methods}

A weak-field expansion is useful in situations where the metric perturbation variables $\Phi$, $\Psi$, $B_i$, 
$h_{ij}$ defined in Eq.~(\ref{e:metric}) are small, as is the case on cosmological scales. In order to get a tractable set of equations,
we retain all terms which are linear in metric
perturbations, but from the quadratic terms we keep only the most relevant ones. These are built from
the scalar perturbations $\Phi$, $\Psi$ only, and contain the highest possible number of spatial
derivatives (two in this case, since all the partial differential equations are second order).
This is enough to ensure that all six metric degrees of freedom are treated correctly at leading order,
even in cases where some of them are strongly suppressed.

\paragraph{Einstein's equations.}
We determine the metric perturbations using the time-time as well as the traceless space-space part
of Einstein's equations given in Eqs.~(\ref{e:modPois}) and~(\ref{e:ij}). The remaining four equations are
redundant, but we can use them as consistency check.
The stress-energy tensor is constructed in the perturbed geometry and may hence also contain some
terms which are linear in the metric perturbations -- these terms have to be taken into account in order
to maintain consistency. For a $\Lambda$CDM cosmology as the one studied here, the stress-energy tensor
is obtained numerically by appropriate particle-to-mesh projections which are ``dressed'' by these geometric
corrections.

We then solve for the metric variables by treating the quadratic terms in Eqs.~(\ref{e:modPois}) and~(\ref{e:ij}) as small perturbations.
To this end, we simply estimate their value using the solutions taken from the previous time step
and move them to the right-hand side. Since the equations are then approximated as linear in the metric
perturbations, we can use Fourier methods to obtain the new solutions. This is convenient in particular since
the gauge conditions can easily be implemented in Fourier space, where they reduce to local projection
operations. One can check that the new solutions are stable by reinserting them into the quadratic terms and
iterating the procedure.

\paragraph{Particle trajectories.}
As dark matter is assumed to be collisionless, the particles move along geodesics. For non-relativistic
velocities $\mathbf{v} = d\mathbf{x}/d\tau$ the geodesic equation reads
\begin{equation}
\label{eq:geodesic}
 \mathbf{v}' + \mathcal{H} \mathbf{v} + \nabla \Psi - \mathcal{H} \mathbf{B} - \mathbf{B}' + \mathbf{v} \times \left(\nabla \times \mathbf{B}\right) = 0 \, ,
\end{equation}
where the $\mathbf{B}$-dependent terms account for frame dragging. The integration of this equation can be simplified
by writing $\mathbf{v} = \tilde{\mathbf{v}} + \mathbf{B}$, which transforms Eq.~(\ref{eq:geodesic}) to
\begin{equation}
 \tilde{\mathbf{v}}' + \mathcal{H} \tilde{\mathbf{v}} + \nabla \Psi + 
 \tilde{v}^i\nabla B_i = 0 \, .
\end{equation}
This new equation actually describes the acceleration as seen in a Gaussian ortho-normal coordinate frame.
Frame dragging, characterized by the last term, has to compete with the ``Newtonian'' force $\nabla \Psi$.
We measure both forces individually in our simulations.

\begin{figure}
\includegraphics[width=0.55\textwidth]{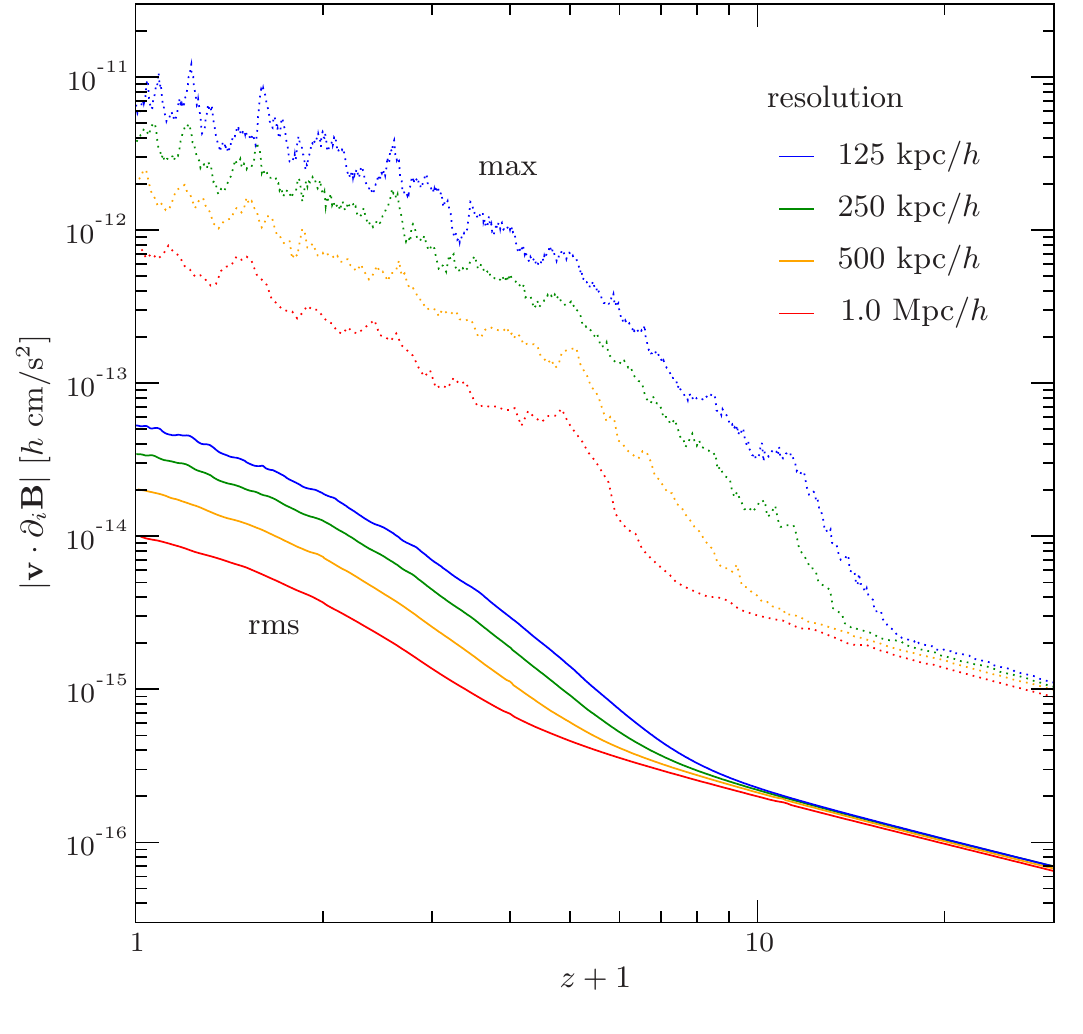}\vspace{-10pt}

\caption{\label{f:framedragging} \small \textbf{Frame dragging.} We show the frame dragging acceleration as a function of redshift $z$ as measured in
simulations of different spatial resolution. The solid lines correspond to mass-weighted rms values, whereas the dotted
lines indicate the maximum value measured in the simulation volume, which was fixed to ($512$ Mpc$/h$)$^3$. The acceleration
is scale dependent and increases with resolution. Although not shown here, we checked that the effects of finite volume and
realisation scatter are less relevant.
}
\end{figure}

Fig.~\ref{f:framedragging} shows the mass-weighted rms value of the frame dragging acceleration
$(\tilde{\mathbf{v}} \cdot \partial_i\mathbf{B})$ for simulations with different spatial resolutions. The simulation volume
was ($512$ Mpc$/h$)$^3$ in all cases, but we used lattice sizes from $512^3$ to $4096^3$ points, reaching a best resolution of
$125$ kpc$/h$. The rms acceleration depends on the probed scale and therefore also on resolution. At our best resolution
we remain slightly above galactic scales. At smaller scales we expect baryonic effects to become important.

\paragraph{Power spectra.}
In Fig.~\ref{f:Pspec} we show the 
power spectra of the metric perturbations which are defined by
\begin{subequations}
\begin{eqnarray}
4\pi k^3\langle \Psi(\bk,z)\Psi(\bk',z)\rangle &=& (2\pi)^3\delta^{(3)}(\bk-\bk')P_\Psi(k,z)\, , \\
4\pi k^3\langle B_i(\bk,z)B_j(\bk',z)\rangle &=& (2\pi)^3\delta^{(3)}(\bk-\bk') P_{ij} P_B(k,z)\, ,\\
4\pi k^3\langle h_{ij}(\bk,z)h_{lm}(\bk',z)\rangle &=& (2\pi)^3\delta^{(3)}(\bk-\bk')M_{ijlm}P_h(k,z) \,. \label{e:Ph}
\end{eqnarray}
\end{subequations}
Here $P_{ij}=\de_{ij}-k_i k_j / k^2$ is the transverse projector and the
spin-2 projection operator is given by $M_{ijlm} = P_{il}P_{jm}+P_{im}P_{jl}-P_{ij}P_{lm}$. These dimensionless power spectra
measure the amplitude square of the metric perturbations at scale $k$ per $\log k$ interval. In order to suppress finite-volume
and resolution effects we only measure scales which are at least $5$ times smaller than the box size and
have a wave number which is at least $5$ times smaller than the Nyqvist frequency.

\paragraph{Initial conditions.}
Initial data is generated using a linear input power spectrum at initial redshift (we start at $z=100$) which can be obtained by
running a Boltzmann code \cite{Lewis:1999bs,Blas:2011rf} for the model. We use \textit{CLASS} with the default cosmological parameters
which describe a $\Lambda$CDM model ($\Omega_b = 0.0483$, $\Omega_m = \Omega_c + \Omega_b = 0.312$, our code currently has no special treatment
for baryonic matter).

\paragraph{Code availability.}
We will release the \texttt{gevolution} code, together with the necessary extensions of the \lftwo library, on a public \textit{Git}
repository (\textit{https://github.com/gevolution-code/gevolution-1.0.git}). We strongly advise interested parties to wait for the fully documented version to become available. Prior to
public release, access to the code can be given upon request, however, we do not offer extensive technical support in this case.

\subsection*{Acknowledgements}

We thank R.~Teyssier and M.~Bruni for discussions. 
This work was supported by the Swiss National Supercomputing Centre (CSCS) under project ID s546.
The numerical simulations were carried out on \textit{Piz Daint} at the CSCS and on
the \textit{Baobab} cluster of the University of Geneva. Financial support was provided by
the Swiss National Science Foundation.

\small
\bibliography{letter}

\end{document}